\documentclass[twocolumn,superscriptaddress]{revtex4} %
\usepackage{graphicx}
\usepackage{epsfig}
\usepackage{amsmath}
\usepackage{amsfonts,amsbsy}
\usepackage{amssymb}
\usepackage{hyperref}
\usepackage{bm}
\usepackage{color}
\usepackage{enumerate}
\usepackage{verbatim}
\usepackage{mathrsfs}

\begin{document}
\title{Relaxation behavior near the first-order phase transition line}
\author{Xiaobing Li}
\affiliation{Key Laboratory of Quark and Lepton Physics (MOE) and Institute of Particle Physics, Central China Normal University, Wuhan 430079, China}
\affiliation{School of Physics and Electronic Engineering, Hubei University of Arts and Sciences, Xiangyang 441053, China}
\author{Ranran Guo}
\affiliation{Key Laboratory of Quark and Lepton Physics (MOE) and Institute of Particle Physics, Central China Normal University, Wuhan 430079, China}
\author{Mingmei Xu}
\email{xumm@ccnu.edu.cn}
\affiliation{Key Laboratory of Quark and Lepton Physics (MOE) and Institute of Particle Physics, Central China Normal University, Wuhan 430079, China}
\author{Jinghua Fu}
\affiliation{Key Laboratory of Quark and Lepton Physics (MOE) and Institute of Particle Physics, Central China Normal University, Wuhan 430079, China}
\author{Lizhu Chen}
\affiliation{School of Physics and Optoelectronic Engineering, Nanjing University of Information Science and Technology, Nanjing 210044, China}
\author{Yu Zhou}
\affiliation{Department of Mathematics, University of California, Los Angeles, California 90095, USA}
\author{Yuanfang Wu}
\email{wuyf@ccnu.edu.cn}
\affiliation{Key Laboratory of Quark and Lepton Physics (MOE) and Institute of Particle Physics, Central China Normal University, Wuhan 430079, China}
\date{\today}
\begin{abstract}
Using the Metropolis algorithm, we simulate the relaxation process of the three-dimensional kinetic Ising model. Starting from a random initial configuration, we first present the average equilibration time across the entire phase boundary. It is observed that the average equilibration time increases significantly as the temperature decreases far from the critical temperature $T_{\rm c}$. The average equilibration time along the first-order phase transition (1st-PT) line exhibits an ultra-slow relaxation. We also investigate the dynamic scaling behavior with system sizes, and find that dynamic scaling holds not only near $T_{\rm c}$, but also at $T\ll T_{\rm c}$. The dynamic exponent at  $T\ll T_{\rm c}$ is larger than that near $T_{\rm c}$. Additionally, we analyze the dynamic scaling of the average autocorrelation time and find that it depends on system size only near $T_{\rm c}$, while it becomes size-independent both above and below $T_{\rm c}$. The extremely slow relaxation dynamics observed near the 1st-PT is attributed to the complex structure of the free energy.
\end{abstract}
 
\maketitle

\section{Introduction}
Understanding relaxation dynamics and scaling behavior is a central goal in studying non-equilibrium systems. While dynamic power-law behavior at critical points (CP) is well-established due to diverging correlation lengths and critical slowing down~\cite{1999-Newman,1977-Hohenberg,1973-Stoll}, relaxation dynamics near first-order phase transitions (1st-PT) remain less explored~\cite{2017-Zhang,2022-Fodor,1972-Stoll,1974-Binder,1992-Tomita,1979-Ikeda,1966-Griffiths,1995-Lee,1992-Berg,1999-Berg,2010-Fischer,2001-Joo,XBLi}. First-order transitions are characterized by metastability and two-phase coexistence, where the transition involves nucleation and domain growth. These processes introduce a more complex relaxation behavior distinct from the critical slowing down observed in second-order transitions.

Dynamic properties such as relaxation rates and multi-time correlation functions depend on the system’s equations of motion. Near CP, relaxation rates approach zero, as shown in conventional~\cite{1954-1,1954-2} and modern theories like mode-coupling and renormalization group approaches, where relaxation rate scales with $\xi^{-z}$, with $z$ representing the dynamic critical exponent~\cite{1977-Hohenberg}.

Two types of relaxation times are commonly discussed: autocorrelation time and non-equilibrium relaxation time~\cite{1973-Stoll}. Autocorrelation time refers to the time scale associated with the transition between one equilibrium state and another. In contrast, non-equilibrium relaxation tracks the system's evolution from a non-equilibrium state to an equilibrium state. However, non-equilibrium relaxation time is specifically defined as the time at which the corresponding relaxation function decreases to $1/{\rm e}$ of its initial value, which does not necessarily indicate that the system has reached true equilibrium~\cite{1974-Binder}.

Near $T_{\rm c}$, both relaxation times follow a power law: 
\begin{equation}
\tau\sim|\varepsilon|^{-z\nu}, \label{tauscaling}
\end{equation}
where $\varepsilon=(T-T_{\rm c})/T_{\rm c}$, $\nu$ is the correlation length exponent, and $z$ is the dynamic exponent. This divergence reflects critical slowing down due to long-range correlations, with numerous studies focusing on accurate evaluations of the dynamic exponent $z$~\cite{1973-Stoll,1993-Mori,2020-Hasenbusch,2023-Roth}. For instance, in Monte Carlo simulations of the 2D kinetic Ising model, both relaxation times exhibit similar exponents~\cite{1973-Stoll}.

Similar slowing-down dynamics also occur in 1st-PT. For example, the autocorrelation time in adsorption-stretching transitions scales with system size~\cite{2017-Zhang}, and in SU(3) Yang-Mills theory, it peaks at 1st-PT temperatures~\cite{2022-Fodor}. Beyond autocorrelation time, the lifetime of metastable states and tunneling time are significant considerations. Metastable states exhibit long lifetimes~\cite{1972-Stoll,1974-Binder,1992-Tomita,1979-Ikeda,1966-Griffiths,1995-Lee}, diverging near the coercive field $H_{\rm c}$. Tunneling time can grow exponentially with system size, termed ``super-critical slowing down", due to large interface energy barriers~\cite{1992-Berg,1999-Berg,2010-Fischer}. However, advanced sampling methods~\cite{1992-Berg,1999-Berg,2010-Fischer} demonstrate power-law scaling instead of exponential growth. Similarly, in the Mott-Hubbard transition, convergence slows near the phase boundary~\cite{2001-Joo}.

In practice, a non-equilibrium relaxation process may start from any initial state and ultimately approach the equilibrium state, where the corresponding observable fluctuates around a stable expectation value. The relaxation time between the initial and final equilibrium states is typically defined as the equilibration time~\cite{1999-Newman}. Understanding equilibration time under varying conditions is crucial for studying non-equilibrium effects and identifying phase boundaries in experimental data~\cite{Mukherjee}. In fast-evolving systems, such as heavy-ion collisions, determining whether equilibrium is achieved is essential, as non-equilibrium observables deviate from their equilibrium values~\cite{XBLi}. However, studies on equilibration time near 1st-PT remain limited, with early research only conceptualizing it~\cite{1999-Newman} without numerical results or detailed analysis.

In our previous studies~\cite{XBLi}, simulations of CP relaxation using the 3D kinetic Ising model~\cite{1963-Glauber,2023-Tikader} showed exponential convergence of the order parameter with time, consistent with Langevin dynamics. The average equilibration time exhibited power-law growth with system size at $T_{\rm c}$, aligning with Model A’s universality class~\cite{1963-Glauber,2020-Hasenbusch,2023-Tikader,1977-Hohenberg}.

In this study, we extend the analysis of equilibration time to 1st-PT. Relaxation dynamics are simulated across the entire phase plane for cubic lattices with periodic boundary conditions. Results reveal that while equilibration and autocorrelation times show similar scaling near $T_{\rm c}$, only the equilibration time diverges at $T\ll T_{\rm c}$. Autocorrelation time exhibits minimal size dependence, whereas equilibration time strongly scales with system size. Additionally, the influence of initial configurations on equilibration time is examined.

The paper is structured as follows: Section II introduces the kinetic Ising model and measurement methods for equilibration and autocorrelation times. Sections III and IV present the simulation results. Section V provides a summary and discussion.

\section{Kinetic Ising model with single-spin flipping dynamics}

The three-dimensional Ising model considers a simple cubic lattice composed of $N=L^3$ spins, where $L$ is called the system size. The total energy of the system with a constant nearest-neighbor interaction $J$ placed in a uniform external field $H$ reads
\begin{equation}
E_{\lbrace s_{i}\rbrace}=-J\sum_{\langle ij\rangle}s_{i}s_{j}-H\sum_{i=1}^N s_{i}, \quad s_{i}=\pm1,
\end{equation}
where $s_i$ denotes the state of the $i$th spin. The per-spin magnetization for each configuration is
\begin{equation}
m=\frac{1}{N} \sum_{i=1}^N s_{i}.
\end{equation} 
It serves as the order parameter of the continuous phase transition at the critical temperature $T_{\rm c}=4.51$~\cite{tc-Ising}. Below $T_{\rm c}$, there is a line of 1st-PT at $H=0$.

In kinetic Ising model, spins are in contact with a heat bath which induces random flips of spins from one state to the other. Only one spin is permitted to flip at once, so that the magnetization is not conserved. The single-spin flipping dynamics with Metropolis algorithm~\cite{Metropolis}, as a local dynamics of Glauber type~\cite{kinetic-Ising}, is suitable for studying nonequilibrium
evolution~\cite{Metropolis-NE1,Metropolis-NE2}. Starting from an initial configuration, a spin is chosen by random for flipping trial. Whether the chosen spin flips depends on the acceptance probability $A({\pmb u}\rightarrow {\pmb v})$, which is given by 
\begin{equation}
A({\pmb u}\rightarrow {\pmb v})=\left\{\begin{array}{ll}
{\rm e}^{-(E_{\pmb v}-E_{\pmb u})/k_{\rm B}T}&\text{if $E_{\pmb v}-E_{\pmb u}>0$},\\1&\text{otherwise.}\end{array}\right .
\end{equation}
${\pmb u}$ and ${\pmb v}$ represent the state of the system before and after flipping this spin. If $A({\pmb u}\rightarrow {\pmb v})=1$, the spin is flipped. If $A({\pmb u}\rightarrow {\pmb v})<1$, a random number $r$ ($0<r<1$) is generated. If $A({\pmb u}\rightarrow {\pmb v})>r$, the spin is flipped, otherwise, the spin keeps its original state. 

The testing of one single spin is called a Monte Carlo step. When $N$ Monte Carlo steps are completed, every spin in the lattice has been tested for flipping and one sweep is completed. In this way, the configuration of the system is updated once a sweep. In the following, time $t$ refers to the number of sweeps. The time evolution of the magnetization is recorded for each evolution process. 

Equilibration time of an evolution process is defined by the number of sweeps required for the order parameter to reach the equilibrium value. Starting from an initial configuration, the magnetization gradually changes until it reaches stable. The stable value is a sign of the equilibrium state. However, after reaching equilibrium, the magnetization still fluctuates around the mean value. The mean value and the standard error of the magnetization are calculated and denoted by $\mu$ and $\sigma$, respectively. In practical operation, the time when the magnetization enters the interval $(\mu-\sigma,\mu+\sigma)$ is recorded as the equilibration time of an evolution process and is denoted by $\tau_{\rm eq}$. 

The real evolution of the magnetization in a single evolution process does not follow the exponential form of a mean-field calculation~\cite{2023-Tikader}, as shown in our earlier paper~\cite{XBLi}. The equilibration time in different evolution processes are different and should be measured event by event. It is necessary to introduce the average equilibration time $\bar{\tau}_{\rm eq}$, representing an average over many evolution processes, i.e. 
\begin{equation}
\bar{\tau}_{\rm eq}=\frac{1}{\mathscr{N}}\sum_{j=1}^{\mathscr{N}}\tau_{\rm eq}^{j},
\end{equation}
where $\mathscr{N}$ is the number of evolution processes. Such defined average equilibrium time indeed characterized the relaxation time of the system near the critical temperature~\cite{XBLi}.

Simulations after the magnetization reaching a stable value is further performed to extract the autocorrelation time. The definition of the autocorrelation function of the magnetization reads
\begin{eqnarray}
\chi(t)=\int \left[ m(t')-\left<m\right>\right]\left[m(t'+t)-\left<m\right> \right]dt'.
\end{eqnarray}
It gives the correlation at two different times, one an interval $t$ later than the other~\cite{1999-Newman}. The autocorrelation function is expected to fall off exponentially as 
\begin{equation}
\chi(t)\sim {\rm e}^{-t/\tau_{\rm auto}}. \label{autocorr}
\end{equation}
The time at which the autocorrelation function $\chi(t)/\chi(0)$ falls to $1/{\rm e}$ is labeled as the autocorrelation time $\tau_{\rm auto}$ for each evolution process. $\bar{\tau}_{\rm auto}$ represents an average over many evolution processes. 

In the following calculation, the absolute value of $m$ is used instead of $m$ because the sign of the magnetization is random at zero field.

\section{Equilibrium time at 1st-PT line and critical point}

A contour plot of the average equilibration time on the $T$-$H$ phase plane for a fixed system size of $L = 60$ is given in Fig.~1. The evolution starts from random initial configurations with all spins pointing randomly up or down. The color scheme ranges from white to red to black, representing the average equilibration time that span from less than a hundred to over four thousand. The phase boundary is depicted by the line $H=0$. In regions far from the phase boundary, the color appears light, indicating a short average equilibration time. A dark-red point emerges around $T_{\rm c}=4.51$, indicating a long average equilibration time, which is consistent with the phenomenon of critical slowing down as expected.

\begin{figure}[tb]
\centering
\includegraphics[width=0.45\textwidth]{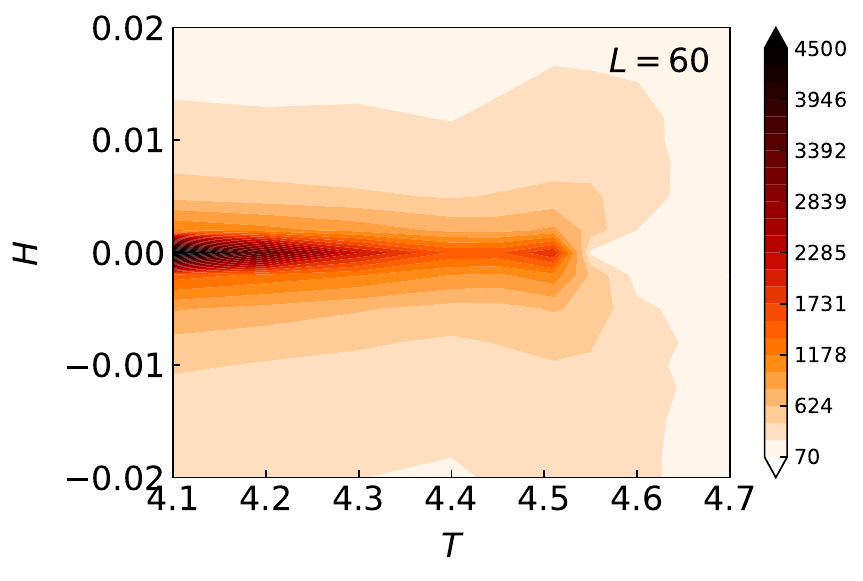}
\caption{A contour plot of the average equilibration time on the phase plane of the 3D kinetic Ising model for $L=60$. The initial configurations are random configurations. }
\end{figure}

Leaving the CP towards the 1st-PT line, the color becomes lighter first and then darker. Further along the 1st-PT line, the color becomes progressively darker as the temperature decreases, eventually turning completely black when the temperature drops below 4.2. Simultaneously, on the low-temperature side, along the direction of the external field, the color transitions rapidly to black near $H=0$. Such a long average equilibration time on and close to the line of 1st-PT indicates that the relaxation process of 1st-PT is considerably slower than that near the CP, i.e. ultra-slow relaxation. 

\begin{figure*}[tb]
\centering
\includegraphics[width=0.8\textwidth]{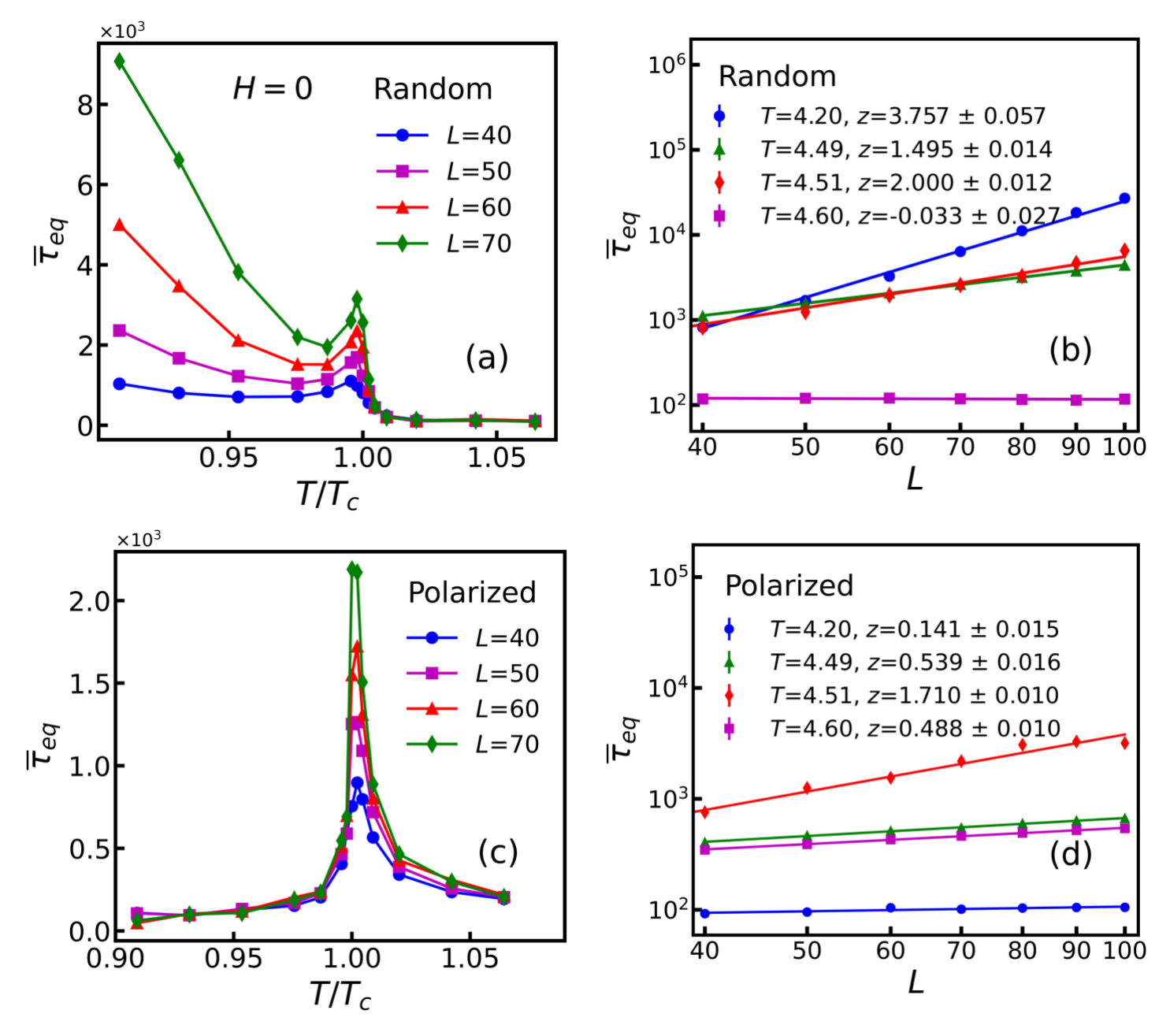}
\caption{(a) The average equilibration time as a function of temperature for four system sizes. Lines are used to guide the eyes. (b) The average equilibration time as a function of system size $L$ for different temperatures in a log-log plot. All points at a given temperature can be well fitted by a line. Figures (a) and (b) show results of the random initial configuration, while figures (c) and (d) are their counterparts of the polarized initial configuration.}
\end{figure*}

To examine the dynamic scaling of the average equilibration time, the dependences on temperature and on system size are shown in Fig.~2. Evolution in upper panels starts from random initial configurations. In Fig.~2(a), near $T/T_{\rm c}=1.00$, the average equilibration time shows a peak. The peak value increases with system size $L$, showing critical slowing down near $T_{\rm c}$. As the temperature decreases from $T_{\rm c}$, the average equilibration time decreases first and then increases. Far below $T/T_{\rm c}\approx 0.98$,  the average equilibration time shows significant size dependence, and the value at some size is even larger than that at $T_{\rm c}$. This size dependence is caused by 1st-PT.  In the following, whenever we say $T\ll T_{\rm c}$, it  means that $T/T_{\rm c}$ is less than 0.98.

The dynamic scaling of the average equilibration time at $T_{\rm c}$ with the system size reads
\begin{equation}
\bar\tau_{\rm eq} \sim L^{z}.\label{scaling-L}
\end{equation}
To cover the size dependence seen below $T/T_{\rm c}\approx 0.98$, the scaling behavior at $T_{\rm c}$ is extrapolated to $T/T_{\rm c}<0.98$. In Fig.~2(b) the scaling is checked from $L=40$ to 100 except $T=4.51$ at which the scaling is checked from $L=20$ to 100 in order to compare with existing data. All points at a given temperature can be well fitted by a line. This shows that the power-law behavior of average equilibrium time, i.e. Eq.~(8), holds across entire phase boundary. Above $T_{\rm c}$ (purple points) the average equilibration time shows almost no dependence on size, while near $T_{\rm c}$ (red points and green points) and far below $T_{\rm c}$ (blue points) the average equilibration time shows strong dependence on size. The exponent at $T_{\rm c}$ is $z=2.000\pm0.012$, which is more precise than the previously reported value of $z=2.06\pm0.03$~\cite{XBLi} due to higher statistics. Both values are consistent with the dynamic universality class of model A~\cite{2020-Hasenbusch}. 

Far below $T_{\rm c}$, e.g. $T=4.2$, we show that the scaling behavior of the average equilibration time with system size still holds. The scaling exponent $z=3.757\pm0.057$ is larger than that of $T_{\rm c}$. As a result, the  average equilibration time at the temperature much lower than $T_{\rm c}$ increases more rapidly with system size than that at $T_{\rm c}$. 

Compared with the CP, the average equilibration time on the line of 1st-PT is longer, its dependence on system size is stronger and the value of the dynamic exponent is larger. All these facts indicate that the slowing down of 1st-PT is more severe than that near the CP.   

To find the impact of initial configurations on the average equilibration time, we repeat the same calculations starting from polarized initial configurations and present the results in Figs.~2(c) and 2(d). Comparing Figs.~2(c) with 2(a), we find that the behavior of the average equilibration time near $T_{\rm c}$ is similar for both initial configurations.  In both cases, when approaching the critical temperature from both sides, the average equilibration time increases, forming a peak. The peak value increases with system size. However, big differences appear far below $T_{\rm c}$. The average equilibration time of polarized initial configuration is short and has almost no size dependence. 

In Fig.~2(d), the dynamic scaling with the system size for polarized initial configurations is shown. Points at a given temperature can be well fitted by a line. The slope at $T_{\rm c}$, i.e. $z=1.710\pm0.010$, is the largest but smaller than that in Fig.~2(b). The slopes of other than $T_{\rm c}$ are all very small. Far below $T_{\rm c}$, e.g. $T=4.20$, the slope is 0.141, indicating almost no size scaling. This is understandable since the polarized initial configurations possess similar structure to the equilibrium states at 1st-PT.   

\begin{figure}[tb]
\centering
\includegraphics[width=0.4\textwidth]{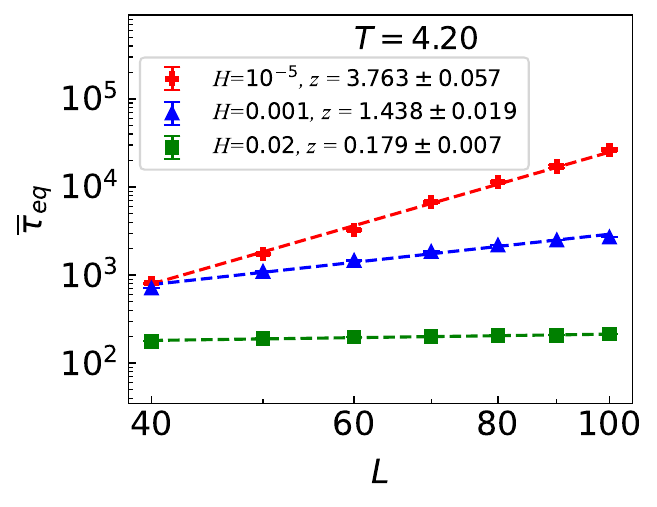}
\caption{The average equilibration time as a function of system size for three values of external field in a log-log plot. Points at a given field can be well fitted by a line. }
\end{figure}

In order to see the dynamic scaling close to the line of 1st-PT,  the average equilibration time for three small fields at $T=4.20$ (far below $T_{\rm c}$) is shown in Fig.~3. Points at a given field can be well fitted by a line. For the smallest field, i.e. $H=10^{-5}$, the exponent $z=3.763\pm0.057$ almost coincides with $z=3.757\pm0.057$ of zero field in Fig.~2(b). When $H$ increases, i.e. leaving the line of 1st-PT further, the exponent $z$ decreases greatly and approximates to zero when $H=0.02$. It shows that the dynamic scaling with the system size still exists when very close to the line of 1st-PT. 

The fact that at zero field and far below $T_{\rm c}$, $\bar\tau_{\rm eq}$ from polarized initial configurations is short while $\bar\tau_{\rm eq}$ from random initial configurations is long, indicates that the equilibrium state of zero field far below $T_{\rm c}$ is hard to reach from random initial configurations. This is understandable from the presence of coexisting states, i.e. the double wells of the free energy. Both upward and downward magnetization are possible equilibrium states, which are the minima of Landau free energy, with $+m_s$ and $-m_{s}$. If starting from upward polarized initial configurations with $m(0)=1$, $m$ is bound to decrease quickly to the right minimum. However, random initial configuration with $m(0)=0$ lies on the top of the barrier between the two minima. The evolution towards two minima is equally probable. Therefore, it is difficult to develop a preferred direction of the spontaneous magnetization. The presence of coexisting states of 1st-PT slows down the relaxation rate, making it difficult to achieve equilibrium. At very small field, the presence of the metastable state, i.e. a shallow minimum of free energy, also slows down the relaxation. Once there is a sizable field which make one of the minima of the free energy disappear, the relaxation to the single minimum becomes fast. In a word, compared to the CP, the relaxation dynamics near 1st-PT are quite different. 

\section{Autocorrelation time near critical point}

For a comparison with $\bar\tau_{\rm eq}$, $\bar\tau_{\rm auto}$ starting from random initial configurations is presented in Fig.~4. As Fig.~4(a) shows, when approaching the critical temperature from both sides, the average autocorrelation time increases, forming a peak. The peak value has size dependence. Larger size, longer $\bar\tau_{\rm auto}$. This is critical slowing down manifested by autocorrelation time. However, except the vicinity of the CP, there is no size dependence at low and high temperature, in contrast to the same cases  of $\bar\tau_{\rm eq}$.   

\begin{figure*}[t]
\centering
\includegraphics[width=0.9\textwidth]{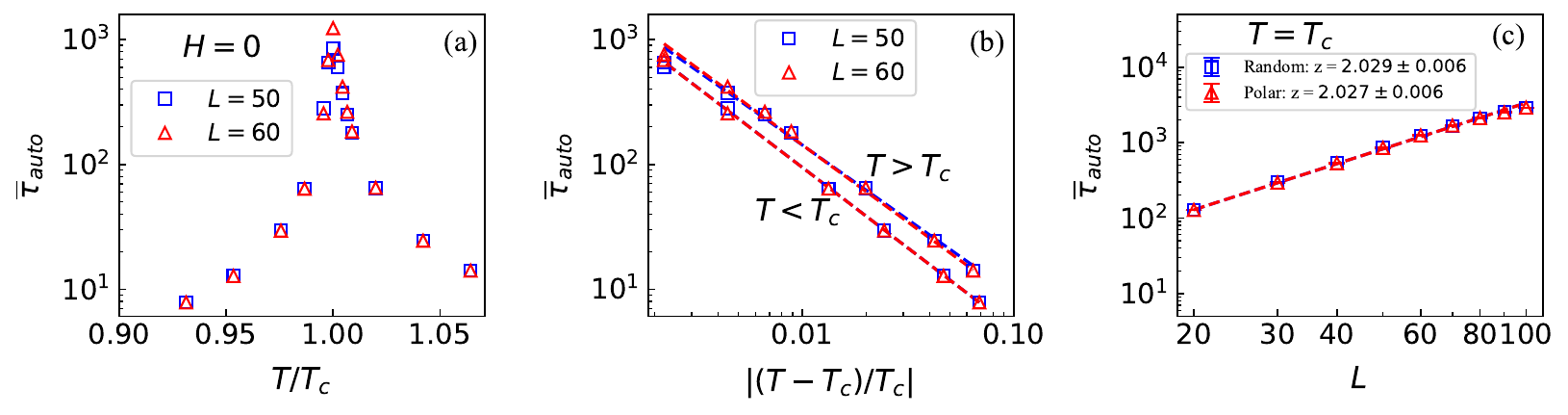}
\caption{(a) The average autocorrelation time as a function of temperature around $T_{\rm c}$ for $L=50$ (blue squares) and 60 (red triangles). (b) The average autocorrelation time as a function of the absolute value of the reduced temperature for $T>T_{\rm c}$ and $T<T_{\rm c}$ with power-law fittings  in a log-log plot. (c) The average autocorrelation time at $T_{\rm c}$ as a function of system size for the random initial configuration (blue squares) and the polarized initial configuration (red triangles) with a power-law fitting in a log-log plot. The exponents of the power law function are shown in the legend.}
\end{figure*}

To show the dynamic scaling of the average autocorrelation time, the dependences on the reduced temperature and on the system size are shown in Figs.~4(b) and 4(c), respectively. In Fig.~4(b), data points for both $T>T_{\rm c}$ and $T<T_{\rm c}$ can be fitted by a power law function in Eq.~(\ref{tauscaling}). This fitting holds up to $|\frac{T-T_{\rm c}}{T_{\rm c}}|\approx0.1$, consistent with Ref.~\cite{1973-Stoll}. 

In Fig.~4(c), the dynamic scaling  of the average autocorrelation time at $T_{\rm c}$ with the system size is checked from $L=20$ to 100. The dynamic exponent $z=2.029\pm 0.006$ is obtained for random initial configurations. The average autocorrelation time of polarized initial configurations are almost equal to that of random configurations. The dynamic exponent extracted for both initial configurations are almost equal, and consistent with other studies~\cite{2020-Hasenbusch}. 

The average autocorrelation time is independent of the initial configuration, whereas the average equilibration time depends on it. This distinction arises from their definitions. The average autocorrelation time is determined after the system has reached equilibrium, and properties measured in equilibrium are generally independent of the path taken to reach it. In contrast, the average equilibration time is inherently linked to the path to equilibrium, which is influenced by the initial configuration. Consequently, the dependence of the average equilibration time on the initial configuration may explain the observed differences in the $z$-value between different types of initial configurations. In conclusion, while the average equilibration time is influenced not only by the system's dynamics but also by its initial configuration, the average autocorrelation time remains unaffected by the initial state.

\section{Summary and discussion}

In this manuscript, we investigate the relaxation behavior near the entire phase boundary of the three-dimensional kinetic Ising model using the Metropolis algorithm. Starting from a random initial configuration, we first present the average equilibration time across the entire phase boundary. It is observed that the average equilibration time increases significantly as the temperature decreases far from the critical temperature $T_{\rm c}$. Compared to the critical slowing down observed near $T_{\rm c}$, the average equilibration time along the 1st-PT line exhibits ultra-slow relaxation.

The dynamic scaling behavior of the average equilibration time as a function of system size is also demonstrated. Our results show that dynamic scaling holds not only near $T_{\rm c}$, but also far below $T_{\rm c}$. The dynamic exponent far below $T_{\rm c}$, e.g. $z = 3.757 \pm 0.057$ at $T=4.2$, is larger than that at $T_{\rm c}$ ($z = 2.000 \pm 0.012$), which is consistent with Model A.

By comparing the average equilibration time with the average autocorrelation time at 1st-PT, we further demonstrate that the average autocorrelation time is independent of the initial configuration or the relaxation process and does not scale with system size. In contrast, the average equilibration time depends not only on the initial configuration but also on the system's dynamical evolution. 

Using the average equilibration time, we derive the dynamic scaling properties not only near the CP, consistent with those obtained using the average autocorrelation time, but also along the 1st-PT line. This demonstrates that the average equilibration time is a powerful observable for studying dynamic scaling. 

The extremely slow relaxation dynamics observed at the 1st-PT is attributed to the complex structure of the free energy, which differs significantly from that near the CP. Near the CP, the free energy has a single minimum, while at the 1st-PT, the free energy has two minima. Along the 1st-PT line ($H = 0$), these two minima are of equal height and correspond to coexisting states. For a random initial configuration with $m(0) = 0$, which lies at the top of the free energy barrier, the evolution towards either minimum is equally probable. This symmetry makes it difficult for the system to develop a preferred direction of spontaneous magnetization, thereby making equilibrium hard to achieve.

Near the 1st-PT line, when the external field is very small, the two minima are no longer equal, with the larger one corresponding to a metastable state. The presence of this metastable state also slows the relaxation process. However, when the external field is sufficiently large, causing one of the free energy minima to disappear, relaxation to the single minimum becomes faster, as observed near the CP.

This feature of the free energy landscape at and near the 1st-PT is universal and not restricted to the three-dimensional Ising model. The observed ultra-slow relaxation at and near the 1st-PT line is expected to be a common characteristic of 1st-PT in general. In addition, the dynamic scaling exponent depends on the relaxation dynamics, such as single-spin flipping Metropolis algorithm, and is therefore model-dependent.

\section*{Acknowledgement}
We are grateful to Dr. Yanhua Zhang for very helpful discussions. This research was funded by the National Key Research and Development Program of China, grant number 2022YFA1604900, and the National Natural Science Foundation of China, grant number 12275102. The numerical simulations have been performed on the GPU cluster in the Nuclear Science Computing Center at Central China Normal University (NSC3). 

\providecommand{\href}[2]{#2}\begingroup\raggedright\endgroup
\end{document}